\documentclass[twocolumn]{emulateapj}

\slugcomment{Accepted for Publication in the Astrophysical Journal Letters}
\submitted{Accepted for Publication in the Astrophysical Journal Letters}

\usepackage{natbib}
\usepackage{apjfonts}
\usepackage{epsf}
\usepackage{mathptmx}

\shorttitle{WeCAPP MACHO Candidates}
\shortauthors{Riffeser et al.}

\newcommand{\E}{\,\mathrm}
\newcommand{\M}{\mathrm}
\newcommand{\Dol}{{D_\mathrm{L}}}
\newcommand{\Dos}{{D_\mathrm{S}}}
\newcommand{\rhol}{{\rho_\mathrm{L}}}
\newcommand{\rhos}{{\rho_\mathrm{S}}}
\newcommand{\RE}{{R_\mathrm{E}}}
\newcommand{\tE}{{t_\mathrm{E}}}
\newcommand{\tfwhm}{{t_\mathrm{fwhm}}}
\newcommand{\ml}{{M}}

\newcommand{\vt}{{v_\mathrm{t}}}

\begin{document}
\title{The Wendelstein Calar Alto Pixellensing Project\altaffilmark{1} 
(WeCAPP): \\ First MACHO Candidates}
\author{Arno Riffeser\altaffilmark{2}, J\"urgen Fliri, Ralf Bender, 
Stella Seitz, and Claus A. G\"ossl}
\affil{Universit\"atssternwarte M\"unchen, Scheinerstrasse 1, 81679 M\"unchen}
\altaffiltext{1}{Based on observations at the Wendelstein 
Observatory of the University Observatory Munich and 
at the German-Spanish Astronomical 
Center, Calar Alto, operated by the Max-Planck-Institut f\"ur Astronomie, 
Heidelberg, jointly with the Spanish National Commission for Astronomy.}
\altaffiltext{2}{\tt arri@usm.uni-muenchen.de}
\begin{abstract}
We report the detection of the first 2 microlensing candidates from
the Wendelstein Calar Alto Pixellensing Project (WeCAPP). Both are
detected with a high signal-to-noise-ratio and were filtered out from
4.5 mill. pixel light curves using a variety of selection criteria. 
Here we only consider well-sampled events with timescales of
$1\E{d} < \tfwhm < 20\E{d}$, high amplitude, and low 
$\chi^2$ of the microlensing fit. The two-color photometry $(R,I)$ 
shows that the events are achromatic and that giant stars with 
colors of $(R-I)\approx 1.1\E{mag}$ in the bulge of M31 have been lensed. 
The magnification factors are 64 and 10 which 
are obtained for typical giant luminosities of $M_I=-2.5\E{mag}$. Both 
lensing events lasted for   only a few days
($t_\M{fwhm}^\M{GL1}=1.7\E{d}$ and  
$t_\M{fwhm}^\M{GL2}= 5.4\E{d}$).  
The event GL1 is likely identical with PA-00-S3 reported by the
POINT-AGAPE project. Our calculations favor in both cases 
the possibility that MACHOs in the halo of M31 caused the 
lensing events. The most probable masses, $0.08\,M_\odot$ for GL1
and $0.02\,M_\odot$ for GL2, are in the range of the brown dwarf limit 
of hydrogen burning. Solar mass objects are a factor of two less likely.
\end{abstract}
\section{INTRODUCTION}
Microlensing experiments are an ideal method to search for dark
objects within and between galaxies.  A large number of microlensing
events have been detected towards the Galactic bulge constraining the
number density of faint stars in this direction \citep{alard99,
afonso99,alcock00a,udalski00}. Towards the LMC only
13-17 microlensing events have been reported so far \citep{alcock00b}.
If all this events are attributed to $0.5\,M_\odot$ MACHOs,
the associated
population of dark objects would contribute up to the 20\% level to the dark
matter content of the Milky Way \citep{alcock00b}. However, both the
relatively large size of the LMC relative to its distance and the
nature of the lenses has cast doubt on this interpretation.  It is
indeed likely that a large fraction of the microlensing events 
towards the LMC are due to self-lensing of stars within the LMC
(\citet{lasserre00,evans00} and references therein).

Studying microlensing events towards M31 allows to separate
self-lensing and halo-lensing in a statistical way, since the optical
depth for halo lensing is larger on the far side of M31. In M31
individual stars can not be resolved and one therefore has to use the
pixellensing technique \citep{crotts92,baillon} to follow the 
variability of sources blended with thousands of other sources within 
the same pixel. First detections of possible microlensing events were 
reported by several pixellensing experiments 
\citep{crotts96,ansari99,auriere01,
paulinhenriksson02,paulinhenriksson03,calchinovati03}.
But since the candidate nature of only 5 of these events 
is convincing, no conclusions concerning the near-far asymmetry or 
the most likely dark matter lensing masses could be drawn yet.

The Wendelstein Calar Alto Pixellensing Project (WeCAPP,
\citealt{arno01}) started in 1997 with test
observations. Since 1999 the bulge of M31 was monitored
continuously during the time of visibility of M31. The analysis 
of our 4 year data will allow not only the identification of very short
duration events (eg., in the 4$^\M{th}$ year data of the combined field
have been taken on 83 \% of possible nights) but also the 
separation of long duration ($<20\E{d}$) microlensing events 
from long periodic variables as Mira stars. For this letter 
we analyzed the short duration events ($\tfwhm<20\E{d}$) 
within one season of Calar Alto data and restricted the detection 
to high-signal-to noise, high-magnification events. We report our 
first 2 microlensing candidates of that type.
\section{OBSERVATIONS AND DATA REDUCTION}
WeCAPP monitors the central region of M31 in a $17.2 \times 17.2\E{arcmin}^2$ 
field with the $1.23\E{m}$ telescope of the Calar
Alto Observatory. In addition, a quarter of this field, pointing
towards the far side of the M31 disk along the SE minor axis, was
observed with the 0.8~m telescope of the Wendelstein Observatory.
The data analysis and candidate selection reported in this letter is
based on the season from June, 23th 2000 to February, 25th 2001 and is
restricted to the Calar Alto data only. During this period, M31 was
observed during 43\% percent of all nights.
Observations were
carried out in $R$ and $I$ filters close to the Kron-Cousins system.
We estimate the systematic error in the $(R-I)$  color to be $\le 0.05\E{mag}$.

We have developed a pipeline based on \citet{arno02} and
\citet{mupipe}, which performs the standard CCD reduction, position
alignment, photometric alignment, stacking of frames, PSF matching
using Optical Image Subtraction \citep{alard98}, and the generation of
difference images.  For the data presented here all data within one
night are coadded, yielding one difference image per night.
The reduction package includes full error propagation for each pixel
through all reduction steps. In this way, all data points are properly
taken into account in the search for variables.
\section{SELECTION CRITERIA}
\begin{deluxetable}{lr}
\tablecaption{Selection criteria
\label{sel}}
\tablehead{
  \colhead{selection criterion}    & \colhead{number} 
}
\tablewidth{0pt}
\startdata
analyzed light curves                                  & 4492250\\
light curves with $>10$ data points                    & 3835407\\
3 successive 3-$\sigma$ in $R$ or $I$                  & 517052\\
$\chi_R^2<1.3$ and $\chi_I^2<1.3$ & 186039\\
1 day $<\tfwhm<20$ days                                & 9497\\
3-$\sigma$ light point inside $\tfwhm$                 & 1829\\
sampling: $\M{side}_1>20\%$, $\M{side}_2>5\%$   & 256 \\
$F_\M{eff}>10\;\M{median_{error}}$ in $R$ and $I$   & 15 \\
candidates                                                & 2  \\
\enddata
\end{deluxetable}
We investigate only pixels which have more than 10 data points in 
$R$ and $I$, which applies for 85\% of the 2K x 2K field. For each pixel we
define a flux baseline by iterative $3\sigma$-clipping of all
outliers with higher flux. All pixels which have at least 3 successive
(positive) 3-$\sigma$ deviations from this baseline are considered as
variables. We fit the microlensing light curve for high-amplification
events \citep{gould} simultaneously to the $R$- and $I$-band pixel light
curves for every variable.  The fit has 6 free parameters:
full-width-half-maximum $\tfwhm$ and the time $t_0$ of
maximum amplification (these 2 parameters are the same for both
filters), amplitude $F_{\M{eff},R}\,$, 
color $F_{\M{eff},I}/F_{\M{eff},R}\,$, and baseline levels $c_R$ and
$c_I$.  Variables with a reduced $\chi_R^2>1.3$ or
$\chi_I^2>1.3$ are discarded. In this way we exclude
light curves that are not achromatic or not symmetric.
We also exclude events with
$\tfwhm>20$ days which can be confused with long periodic
variables like Mira stars, as long as only one season of data is
investigated.  In addition, all candidates which do not have at least
one significant data point (3 $\sigma$ deviation from the baseline)
within $\tfwhm$ of the time of maximum amplification are
rejected.
We further define the sampling quality for the falling and rising
parts of each light curve within $(t_0-15\E{d},t_0)$ and
$(t_0,t_0+15\E{d})$: 
within these time intervals we require a sampling of the area under
the light curve of at least 20\% on one side and of at least 5\% on
the other side (Table \ref{sel}).

\begin{figure}
\plotone{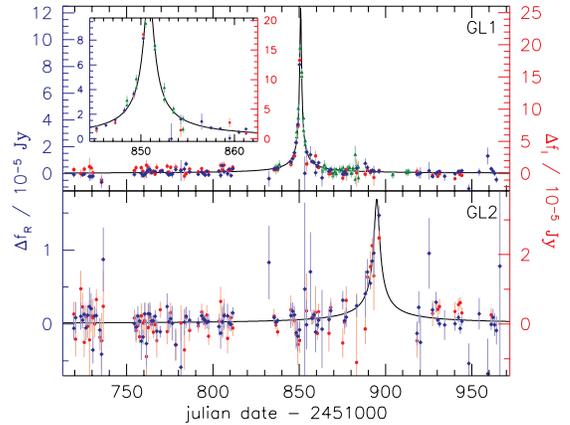}
\caption{
Light curves  of WeCAPP-GL1 and WeCAPP-GL2.
The I-band light curve (red symbols, right axis) 
has been scaled to the $R$-band light curve (blue symbols, left axis). 
The scaling factors were derived from the lensing fit (black curve) and 
correspond to a color $(R-I)$ of 1.05 for GL1 and 1.08 for GL2.
In addition we show the $r'$ and $i'$ data from the POINT-AGAPE PA-00-S3 
event (green symbols) scaled to our data. \label{gl}}
\end{figure}

Here, we present only the two microlensing candidates which have
amplitudes 10 times larger than the median error of the light curve
(see Figure \ref{gl}). 
Both candidates fit perfectly to a symmetric microlensing light curve. 
Ruling out systematic offsets for the points and errors on the trailing side
of GL2 (which is strongly proved by the 6 single images of that night 
in each filter) a non microlensing light curve of a variable 
source hardly fits the data points of GL2.
Both microlensing candidates are detected in several pixels
($11$ for GL1 and $4$ for GL2) inside the PSF of the position of the
lensed object.  This explains the reduction from 15 events to 2 events
in the last line of Table \ref{sel}. The amplification light curves were
obtained by calculating the total flux within the PSF area of each
microlensing event.  
For both candidates, the selection criteria
exclude variable stars like Miras, novae or dwarf novae.
Extracting lensing events with less good time sampling, lower amplitude, or
events located close to other variables requires refined selection
criteria. These will be discussed in a future paper which will also
include a test of the detection efficiency and false detection rate
with Monte Carlo simulations.
\section{MICROLENSING CANDIDATES}
\begin{deluxetable}{cccc}
\tablecaption{Positions and parameters of the microlensing candidates 
\label{pos}}
\tablehead{
 Name & \colhead{WeCAPP-GL1} & \colhead{GL1 \& PA-00-S3 \tablenotemark{1}}& 
\colhead{WeCAPP-GL2}
}
\tablewidth{0pt}
\startdata
 $\alpha_{2000}$ & $00^h 42^m30 \fs 3$  & $00^h 42^m30 \fs 3$ 
& $00^h 42^m 32 \fs 8$ \\
 $\delta_{2000}$ & $+41\arcdeg 13\arcmin 00 \farcs 8$ 
& $+41\arcdeg 13\arcmin 00 \farcs 8$ & $+41\arcdeg 19\arcmin 56\farcs 5$ \\
 $t_0$ [JD]   & $2451850.80\pm 0.13$  & $2451850.84\pm 0.02$     
& $2451894.77\pm 0.21$ \\
 $\tfwhm$ [d]     & $1.38\pm 0.53$    & $1.65\pm 0.10$     & $5.41\pm 2.49$ \\
 $F_{\M{eff},R}$  &   $13.4 \pm 5.4$  &   $12.4 \pm 0.6$   & $1.7 \pm 0.5$ \\ 
 $F_{\M{eff},I}$  &   $28.0 \pm 11.2$ &   $25.7 \pm 1.5$  & $3.6 \pm 1.1$ \\
 $(R-I)$	  & $1.05\pm 0.08\pm 0.05$ & $1.05\pm 0.08\pm 0.05$   
& $1.08\pm 0.24\pm 0.05$ \\
 ${\chi}^2$ & 1.23       & 1.22       & 1.02 \\
 \enddata
\tablenotetext{1}{derived from a fit to the total set of data points 
(WeCAPP \& POINT-AGAPE)}
\end{deluxetable}
The parameters of both lensing candidates are summarized in Table \ref{pos}.
Their light curves are shown in Figure \ref{gl}. GL1, the highest S/N 
lensing event candidate in our sample, lies $4\fm1$
to the SW of the nucleus of M31. GL2 is $4\fm4$ to the NW of the
nucleus. 
Our data have been astrometrically calibrated using bright foreground
stars observed with HST by \cite{jablonka} and ground based
observations by \cite{magnier}. 
Our two calibrations agree within 
0.5$\arcsec$ in declination and 0.7$\arcsec$ in right ascension, well
consistent with the astrometric accuracy of 0.8$\arcsec$ to 
1.0$\arcsec$ of the Magnier et al.\ catalog. 
After we had detected GL1 and GL2, we cross-checked with
events reported by the POINT-AGAPE survey for the same period of time
and the same field in M31 \citep{paulinhenriksson03}. It appears that GL1 is 
likely identical with PA-00-S3 which occurred at the same time 
(Figure \ref{gl}).
Because POINT-AGAPE did not
provide a flux calibration of their data, we had to assume a scaling
factor for the amplitude. The zero point in time was not adjusted.
The data points from WeCAPP and POINT-AGAPE complement each other
nicely and make GL1 the best pixellensing event found so far in M31.
GL2 also falls in the observing period covered by POINT-AGAPE but
their time sampling around the event is poor. This may be the reason 
why GL2 was not detected.

The parameters of the lensing fit are degenerate for  high 
magnifications \citep{gould}, i.e. for amplitudes $A_0 \gg 1$
which correspond to  impact angles
much smaller than the Einstein angle $\theta_\M{E}$.
In this case,  we obtain for the light curve:
\begin{displaymath}
F_0\,[A(t)-1] \longrightarrow
F_\M{eff}\,\left[\frac{12}{\tfwhm^2}(t-t_0)^2+1\right]^{-1/2}
\end{displaymath}
with intrinsic  flux of the source $F_0$
and $F_\M{eff}=F_0/u_0\approx F_0 A_0$, where $u_0$ is the impact angle 
in units of the Einstein angle.

In order to obtain $A_0$ we evidently need to know the source flux
$F_0$.  We can get rough constraints by considering the color of the
light curves which due to our selection criteria are achromatic.  We
obtain $(R-I)_\M{GL1}=1.05$ and
$(R-I)_\M{GL2}=1.08$.
For main sequence stars this 
converts\footnote{transformed on the observational plane by \cite{maraston98}} 
\citep{cassisi} into absolute magnitudes of
$M_I\approx 8\E{mag}$.
If the sources are giants, then the magnitudes are 
$M_I=-2.5\E{mag}$ with a minimum of $-1.9$ and
maximum of $-3.7$ \citep{bessell,grillmair}.

We now derive the lensing parameters as follows:
\begin{displaymath}
u_0=\frac{F_{0,I}}{F_{\M{eff},I}} \;\longrightarrow\;
\tE=\frac{\tfwhm}{u_0\sqrt{12}}
\;\longrightarrow\;
M_\M{lens}=\frac{(\vt \tE) ^2 c^2 \Dos}{4G
  \Dol(\Dos-\Dol)}
\end{displaymath}
where $\tE$ is the Einstein timescale, $v_t$ is the transverse velocity 
between source and lens, and $D_L$ and $D_S$ are the angular distances to 
the lens and source.
\begin{deluxetable}{rcc}
\tablecaption{Amplifications for different source star luminosities
\label{amplifications}}
\tablehead{\colhead{$M_I$} & \colhead{$A_0$} & \colhead{$t_E$[d]}}
\tablewidth{0pt}
\startdata
\bf GL1 \\
 -1.9 &    120 &    $47.4 \pm    18.1$  \\
 -2.5 &     69 &    $27.2 \pm    10.4$  \\
 -3.7 &     24 &     $9.0 \pm     3.5$  \\
  7.7 &  $8.2\cdot 10^5$ & $(3.3\pm 1.3)\cdot 10^5$ \\
\bf GL1 & \bf \& \,PA-00-S3 \\
 -1.9 &    110 &    $52.1 \pm     3.2$   \\
 -2.5 &     64 &    $30.0 \pm     1.8$   \\
 -3.7 &     22 &   $  9.9 \pm     0.6$   \\
 7.7 & $7.6\cdot 10^5$ & $(3.6\pm 0.2)\cdot 10^5$ \\
\bf GL2 \\
 -1.9 &     16 &    $23.9 \pm    11.0$  \\
 -2.5 &     10 &    $13.8 \pm     6.3$  \\
 -3.7 &      4 &    $ 4.6 \pm     2.1$  \\
  8.3 &  $1.8\cdot 10^5$ & $(2.9\pm 1.3)\cdot 10^5$ \\
\enddata
\end{deluxetable}

To estimate plausible lens masses the Einstein timescales $\tE$ are
 calculated for fixed luminosities of possible source stars (see Table
\ref{amplifications}). Note that the errors in $\tE$ reflect 
the accuracy of the determination of $\tfwhm$ in the degenerate 
Gould fit only, and do not account for the systematic uncertainties due to
the unknown luminosities of the sources. 
If the source is a main sequence star, we need very high magnifications:
typically $A_\M{0}= 8\cdot10^5$ and $A_\M{0}= 2\cdot 10^5$
for GL1 and GL2, respectively. The corresponding lens masses 
(for $v_t=210\E{km/s}$, 
$\Dol=768\E{kpc}$,
$\Dos=770\E{kpc}$)
are $M\approx 10^8\,M_\odot$, an implausibly large value. If the source is
a giant, the required magnifications are reduced to
$A_\M{0,GL1}=64$ and $A_\M{0,GL2}=10$ (see Table
\ref{amplifications}); the typical self-lensing
masses become $M=0.8\,M_\odot$ and $M=0.2\,M_\odot$, i.e. they are
typical for low mass stars.

Assuming the source to be a red giant with $M_I=-2.5\E{mag}$
we calculate the probability $p(M,t_E)$ that a microlensing event of 
observed timescale $t_E$ can be produced by a lens of the mass $M$.
Following the calculations of 
\cite{jetzer94a} eq. (8) and \cite{jetzer94b} eq. (11) we get
\begin{displaymath}
\begin{array}{l}
  \displaystyle
  p(M,\tE)\sim \,\xi(\ml) 
  \int \rhos(\Dos) \int \rhol(\Dol)  
  \,{f\left(\frac{\RE}{\tE}\right)} 
  \,\frac{\RE^3}{\tE^3} \,d\Dol \,d\Dos 
\end{array}
\end{displaymath}
with the mass function (MF) $\xi(\ml)$, the sources density $\rhos(\Dos)$, 
the lenses density $\rhol(\Dol)$,
the velocity distribution $f(\vt)$ and 
the Einstein radius $\RE(\Dol,\ml,\Dos)$.

The distribution of matter in the central part of M31 is based on the 
bulge model of \cite{kent}. The disk is modeled with a radial scale length 
of 6.4 kpc and an exponential shape, and with a vertical scale length of 
0.3 kpc and a $\M{sech}^2$-shape.
The halo is modeled as an isothermal sphere with a core radius of 
$r_\M{c}=2\E{kpc}$. 
The velocity distribution was calculated from a Maxwellian halo 
bulge and disk
velocity distribution with an additional rotation for bulge 
and disk \citep{kerins01}.

For the bulge lenses we take the IMF as derived for the galactic
bulge $\xi \sim M^{-1.33}$ \citep{zoccali}.
For the disk population we adopt a Gould IMF $\xi \sim M^{-2.21}$ 
with a flattening $\xi \sim M^{-0.56}$ below $0.59\,M_\odot$ \citep{gould97}.
Both are cut at the lower end at the hydrogen burning limit of 
$0.08\,M_\odot$. At the upper end the bulge MF is cut at the main sequence 
turn-off $0.95\,M_\odot$ (Maraston, priv. com.)
and the disk MF at $10\,M_\odot$.
The IMF for the potential MACHO population residing in the 
halo of M31 is of course unknown. We therefore calculate the probability 
distribution for halos consisting of one mass only, i.e.~taking 
$\delta$-function IMF's centered on the lens mass 
$\xi = \delta(M-M_\M{lens})/M_\M{lens}$. Moreover
we assume that the whole dark halo of M31 consists of MACHOs.
Lensing by Galactic halo objects has an order of magnitude smaller 
optical depth and is therefore neglected in our considerations.

\begin{figure}
\plotone{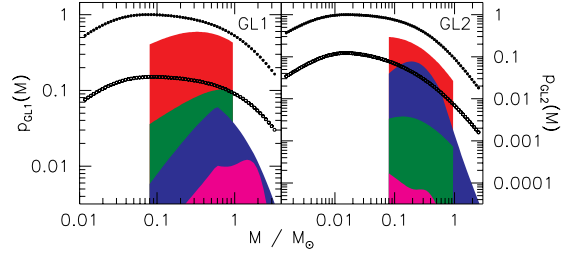}
\caption{Mass probability for GL1 (left panel)
and GL2 (right panel) 
for lens-source configurations: 
halo-bulge (filled circles), halo-disk (open circles), 
bulge-bulge (red), bulge-disk (green), 
disk-bulge (blue), disk-disk (magenta).
The maximum of each curve is scaled to reflect
the total probability of a respective lens-source event relative 
to the case of a halo-bulge lensing event with the most probable
MACHO mass. For example in case of GL1 the probability for bulge-bulge lensing 
relative to halo-bulge lensing with $0.08\,M_\odot$ lenses 
becomes 0.6 (maximum of red curve). A halo consisting of 
$0.014\,M_\odot$ MACHOs would have the same probability 
as bulge-bulge lensing.
Note that the shapes of the distributions for bulge and disk lenses 
are strongly affected by the used mass function $\xi(M)$. 
\label{masses}}
\end{figure}

The results are  shown in Figure \ref{masses}.
For M31 halo lenses the most probable masses are $0.08\,M_\odot$ for GL1
and $0.02\,M_\odot$ for GL2. In the case of self-lensing the most probable 
masses are about a factor of 4 bigger.
Taking the most likely halo lens masses, the ratio of the probabilities 
that the lenses are part of the dark halo 
or the stellar content $p_\M{halo}/(p_\M{bulge}+p_\M{disk})$ is 1.6 for 
GL1 and 3.3 for GL2. We conclude therefore that it is likely that lenses 
residing in the halo of M31 caused the events in both cases.
\section{DISCUSSION and OUTLOOK}
We presented the first two high ($S/N$), short timescale
microlensing events from WeCAPP. GL1 is likely identical to PA-00-S3
found by POINT-AGAPE. 
Combining the data from AGAPE with ours shows that
the error bar of the derived Einstein time scale becomes smaller by a
factor of 5 compared to the individual error bar.
This demonstrates the importance of a good time
sampling of the events.
We derived the colors of the lensed
stars, the amplification factors and likely lens masses for both bulge/disk
self-lensing and MACHO lensing. We showed that red giants are the
likely source objects, while main sequence stars are highly
implausible. 

Self-lensing in the bulge can only be separated from MACHO lensing
statistically.  Halo-lensing events show a spatial asymmetry because
the optical depth for lensing events is higher for stars on the far
side of M31 than on its near side \citep{crotts92,theory}. In contrast,
bulge self-lensing is symmetric.

The bulge self-lensing hypothesis yields lensing stars at or below the
the main sequence turn-off of the M31 bulge. On the other hand, if the 
lensing events are caused by MACHOs, their masses are typically
very low, most probable below $0.1\,M_\odot$. Masses in the range of
$0.5\,M_\odot$ to $1\,M_\odot$ are more unlikely.

So far, we have  analyzed one observing season and restricted the 
lensing search to short-time, high-amplification events in order to avoid 
confusion with variable stars. The whole WeCAPP dataset
will allow us to identify all
variables and thus will enable a search for lower amplitude and longer 
duration microlensing events. 

Decreasing the amplitude threshold will increase the detected
rate of events in two ways. As the event rate is proportional 
to the inverse of the minimum required magnification $A_{0,\M{min}}$
in the pixellensing regime we expect to detect more lensed giants.
On the other hand lowering the amplification threshold could
make it possible to detect also highly amplified main 
sequence stars \citep{hangould} which exceed the evolved stars 
in the bulge of M31 by a factor of more than a hundred. How many 
more lensing events will be detected depends on the mass function 
of the lenses but we can expect at least a factor of a few \citep{theory}.

Finally, the effects of time sampling and noise properties 
of our sample on the detectability of lensing events have to be 
taken into account. Results of the modeling of these effects for 
events of different durations and amplitudes using Monte-Carlo 
simulations will be presented in a future publication. 
With the full dataset we expect therefore to increase the 
number of lensing events to detect the predicted asymmetry 
of MACHO lensing or to rule out a significant MACHO population in 
the halo of M31.
\acknowledgments
This work was supported by the {\em SFB 375
 Astro-Particle-Physics} of the Deutsche Forschungsgemeinschaft.
\end{document}